\documentclass[%
preprint,
 amsmath,amssymb,
 aps,
prb,
]{revtex4-1}

\usepackage{graphicx}
\usepackage{dcolumn}
\usepackage{bm}
\usepackage[usenames,dvipsnames]{xcolor}
\usepackage[version=3]{mhchem}
\usepackage{float}
\floatstyle{plaintop}
\restylefloat{table}
\usepackage{hyperref}

\begin{document}


\title{Critical size limits for collinear and spin spiral magnetism in \ce{CoCr2O4}}

\author{D. Z\'{a}kutn\'{a}$^{1,2}$}
\author{A. Alemayehu$^{3}$}
\author{J. Vl\v{c}ek$^{4}$}
\author{K. Nemkovski$^{5}$}
\author{C. P. Grams$^{6}$} 
\author{D. Ni\v{z}\v{n}ansk\'{y}$^{3}$}\thanks{Deceased}
\author {D. Honecker$^1$}\thanks{now at: Physics and Materials Science Research Unit, University of Luxembourg, 162A Avenue de la Faıencerie, L-1511 Luxembourg, Grand Duchy of Luxembourg}
\author{S. Disch$^2$} 
\thanks{corresponding author: sabrina.disch@uni-koeln.de}
\affiliation{$^1$Institut Laue-Langevin, 71 Avenue des Martyrs, F-38042 Grenoble, France}
\affiliation{$^2$Department f{\"u}r Chemie, Universit{\"a}t zu K{\"o}ln, Luxemburger Strasse 116, 50939 K{\"o}ln, Germany}
\affiliation{$^3$Department of Inorganic Chemistry, Faculty of Science, Charles University in Prague, Hlavova 2030/8, 12843 Prague 2, Czech Republic}
\affiliation{$^4$Department of Physics and Measurements, University of Chemistry and Technology Prague, Technicka 5, 16628 Prague, Czech Republic}
\affiliation{$^5$Forschungszentrum J{\"u}lich GmbH, J{\"u}lich Centre for Neutron Science (JCNS) at Heinz Maier-Leibnitz Zentrum (MLZ), Lichtenbergstr. 1, 85748 Garching, Germany}
\affiliation{$^6$ II. Physikalisches Institut, Universit{\"a}t zu K{\"o}ln, Z{\"u}lpicher Str. 77, 50937, K{\"o}ln, Germany}

\date{\today}

\begin{abstract}
The multiferroic behavior of \ce{CoCr2O4} results from the appearance of conical spin-spiral magnetic ordering, which induces electric polarization. The magnetic ground state has a
complex size dependent behavior, which collapses when reaching a critical particle size.
Here, the magnetic phase stability of \ce{CoCr2O4} in the size range of 3.6 - 14.0\,nm is presented in detail using the combination of neutron diffraction with XYZ polarization analysis and macroscopic magnetization measurements. We establish critical coherent domain sizes for the formation of the spin spiral and ferrimagnetic structure and reveal the evolution of the incommensurate spin spiral vector with particle size. We further confirm the presence of ferroelectric polarization in the spin spiral phase for nanocrystalline \ce{CoCr2O4}.
\end{abstract}

\maketitle


The coupling of two ferroic properties, known as multiferroism is widely studied due to its applicability in spintronics\cite{Fiebig2016}. Many materials, such as \ce{RMnO3} and \ce{RMn2O5} (R = Dy, Tb or Ho) have ferroelectric properties emerging in an antiferromagnetic state\cite{Kimura2003,Goto2004,Bousquet2016,Higashiyama2004}. Multiferroics, for which the ferroelectricity arises from a spiral magnetic magnetic structure are in general considered as ferroelectrics showing the strongest sensitivity to magnetic field\cite{Mostovoy2006}. In the case of \ce{CoCr2O4}, ferroelectric polarization arises from the spin-spiral structure, which is formed due to the strong direct exchange interaction in the B spinel site\cite{Cheong2007}. Thus, \ce{CoCr2O4} is a multiferroic material with both spontaneous magnetization and electric polarization of spin spiral origin\cite{Yamasaki2006,Kim2009,Chang2009}. The magnetic properties of nanocrystalline \ce{CoCr2O4} change significantly from its bulk counterpart, for instance no lock-in transition $T_\mathrm{lock}$ to the ground state is found\cite{Zakutna2014,Zakutna2016,Zakutna2018,Rath2011,Dutta2009,Kumar2009,Tian2015,Galivarapu2016,GalivarapuRSC2016}. Moreover, it is difficult to deduce clearly the spin spiral transition temperature $T_\mathrm{s}$ and the presence of the associated spin spiral magnetic ordering from volume averaged magnetization measurements\cite{Zakutna2014}. Few studies employed polarized neutron diffraction measurements, where the appearance of magnetic satellite reflections is a clear indicator of the spin spiral magnetic ordering in \ce{CoCr2O4}\cite{Zakutna2018,Galivarapu2016,GalivarapuRSC2016} and Fe-doped \ce{CoCr2O4}\cite{Kumar2016}. 
A strong influence of the particle size on the magnetic properties is suggested\cite{Zakutna2018}: nanoparticles (NPs) with an average grain size of 22\,nm reveal a clear spin spiral configuration below $T_\mathrm{s}$ = 27\,K, similar to the bulk material, whereas small NPs of 3\,nm size exhibit a collective cluster glass behavior.\\
Here we present a detailed magnetic phase diagram of \ce{CoCr2O4} for the size range of 3.6 - 14.0\,nm obtained by combining polarized neutron diffraction with macroscopic magnetization measurements. The grain-size and temperature dependent boundaries of the magnetic phases are deduced from the fundamental reflection for collinear magnetic ordering and magnetic satellites of spin spiral ordering. Next to the critical size limits for both collinear and spin spiral magnetism, we confirm multiferroic properties in the spin spiral phase through ferroelectric polarization.\\

\begin{figure}[htpb]
	\centering
	\includegraphics[width=0.4\textwidth]{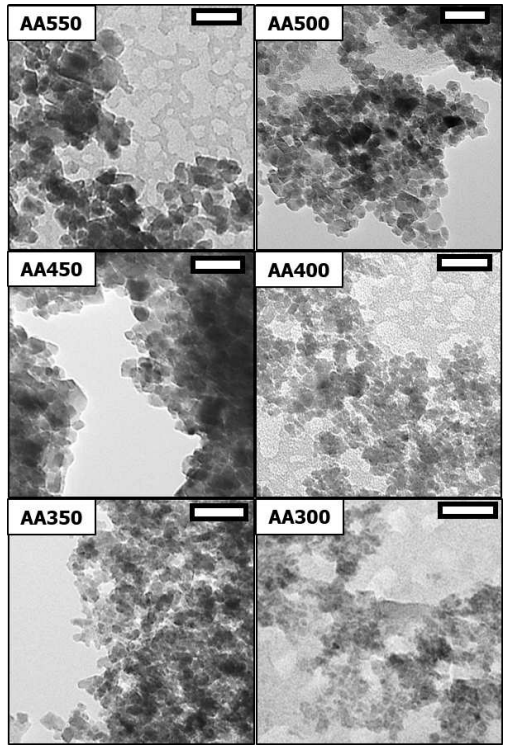}
	\caption{BF TEM micrographs of cobalt chromite samples. (scale bars: 50\,nm).}
	\label{fig:XRD}
\end{figure}

Cobalt chromite NPs with tunable particle size and good crystallinity are obtained by temperature dependent annealing of the amorphous \ce{CoCr2O4} precursor gained from hydrothermal synthesis (for more information see\cite{SupInf}). The cobalt chromite NPs have nearly spherical morphology as confirmed by TEM (\textbf{Fig. \ref{fig:XRD}}). The samples are labeled as AA[annealing temperature in\,$^{\circ}$C]. PXRD data (\textbf{Fig. S 1})\cite{SupInf} were refined according to the spinel structure of \ce{CoCr2O4} ($F d \bar{3} m$ space group) using Le-Bail analysis and confirm phase purity for all samples. A small impurity of NaCl was detected for the sample AA350 and removed before the neutron scattering experiment by a washing step. The structurally coherent particle sizes from PXRD are in good general agreement but slightly smaller than the particle sizes obtained by TEM analysis (\textbf{Fig SI 2}\cite{SupInf}) indicating structural disorder near the particle surface. The size distribution of the NPs is reasonably narrow and in the range of $\sigma_\mathrm{log}$ = 0.1 - 0.2 (\textbf{Table SI II}\cite{SupInf}). The nanoparticle size increases with annealing temperature, and simultaneously the lattice parameter slowly approaches the bulk value (Fig. S2\cite{SupInf}).

\begin{figure}[htpb]
	\centering
	\includegraphics[width=0.8\columnwidth]{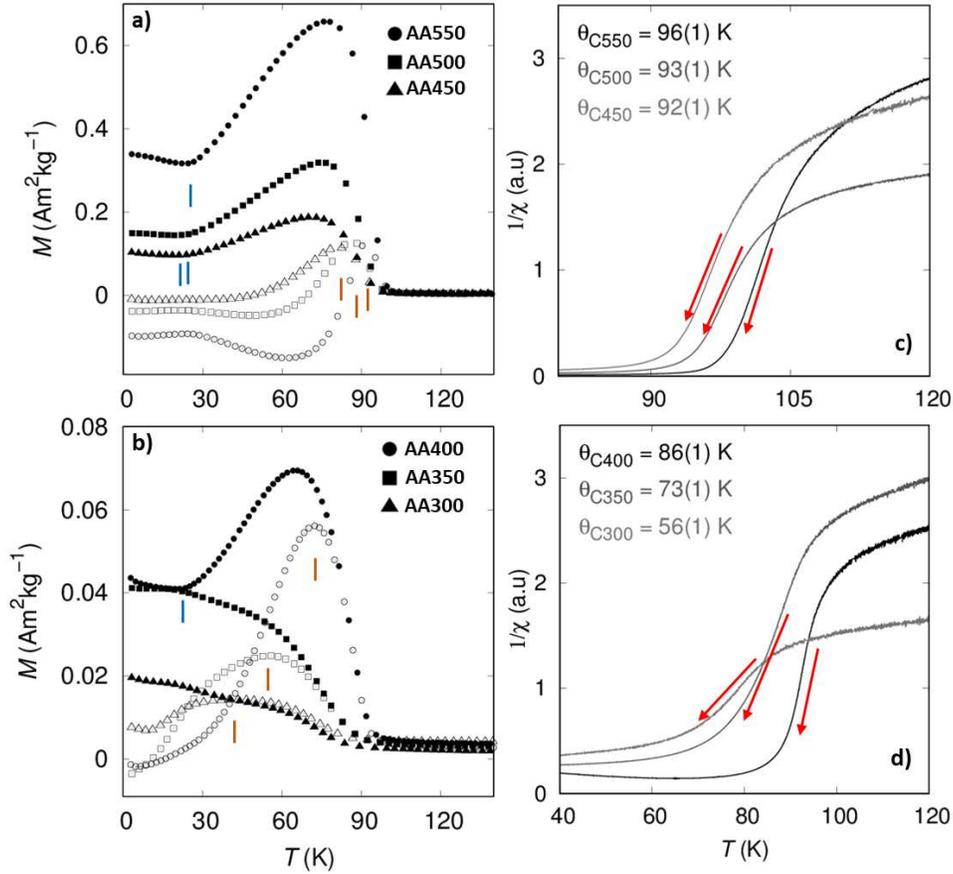}
	\caption{\textbf{a-b)} ZFC/FC (full/open symbols) magnetization measurements of cobalt chromite NPs with different particle sizes. Blue and brown vertical lines correspond to the transition temperatures $T_\mathrm{s}$ and $T_\mathrm{b}$, respectively. \textbf{c-d)} the temperature dependence of the inverse susceptibility. Arrows indicate the linear extrapolation in the paramagnetic regime to obtain the Curie temperature.}
	\label{fig:ZFC}
\end{figure}

The field cooled (FC) and zero-field cooled (ZFC) magnetization measurements (\textbf{Fig. \ref{fig:ZFC} a-b)}) reveal three distinct magnetic phase transitions. The Curie temperature $\theta_\mathrm{C}$ is determined by linear extrapolation of the high temperature magnetic susceptibility (\textbf{Fig. \ref{fig:ZFC} c-d)}). The blocking temperature $T_\mathrm{b}$, corresponding to the transition from superparamagnetic to the ferrimagnetic blocked state, is obtained from the maximum of the ZFC curve (\textbf{Table S III}\cite{SupInf}). AC susceptometry reveals only a slight temperature dependence of the blocking temperature with AC frequency (\textbf{Fig. S 3})\cite{SupInf} indicating a well-ordered collinear magnetic state. Further decreasing the temperature, a shallow minimum of the FC magnetization is observed indicating the transition temperature $T_\mathrm{s}$ to spin spiral magnetic ordering. The so-called lock-in transition which is reported for bulk material ($T_\mathrm{lock} \approx 15$\,K) is not observed for any of our nanocrystalline samples. Comparing the ZFC/FC curves, a shift of Curie, blocking and spin spiral transition temperatures towards smaller temperatures is clearly visible with decreasing particle size. For a coherent domain size of less than $d_\mathrm{XRD} =$ 6.4\,nm (AA400, AA350, AA300), the minima of the FC curve corresponding to the spin spiral magnetic ordering are strongly suppressed.
The magnetic phase transitions are also reflected in isothermal magnetization measurements (\textbf{Fig. S 4}\cite{SupInf}). Above the Curie temperature, a linear magnetization dependence is observed corresponding to the paramagnetic state. The SPM state is reflected by a Langevin shaped magnetization curve. Below $T_\mathrm{b}$, hysteresis appears related to the blocked ferrimagnetic order. At the base temperature of 2\,K the magnetization is non-saturated up to high magnetic fields of 6\,T, indicating spin disorder effects in the NPs. This is in line with the significantly different particle sizes observed by PXRD and TEM, where the crystalline part is smaller than the particle size. The structural disorder, related to the lack of crystallinity at the particle surface, is correlated with the decrease of the spontaneous magnetization observed with decreasing particle diameter (\textbf{Table S III}\cite{SupInf}).

Neutron diffraction has been employed to investigate the collinear and non-collinear magnetic phases and to resolve the particle size dependence of the magnetic phase transitions in more detail. Polarized neutron diffraction with XYZ polarization analysis gives the opportunity to clearly separate the spin-incoherent, nuclear-coherent and magnetic scattering contributions (see \textbf{Fig S 5} for the complete sample set at 3.5\,K). The spin-incoherent diffuse scattering contribution varies with increasing annealing temperature and particle size. We attribute this observation to the different amount of oleic acid surfactant in the sample due to the varying surface to volume ratio and decomposition of the oleic acid starting at $\approx$ 350\,$^\circ$C\cite{Zakutna2014}.  

\begin{figure}[htpb]
	\centering
	\includegraphics[width=0.8\columnwidth]{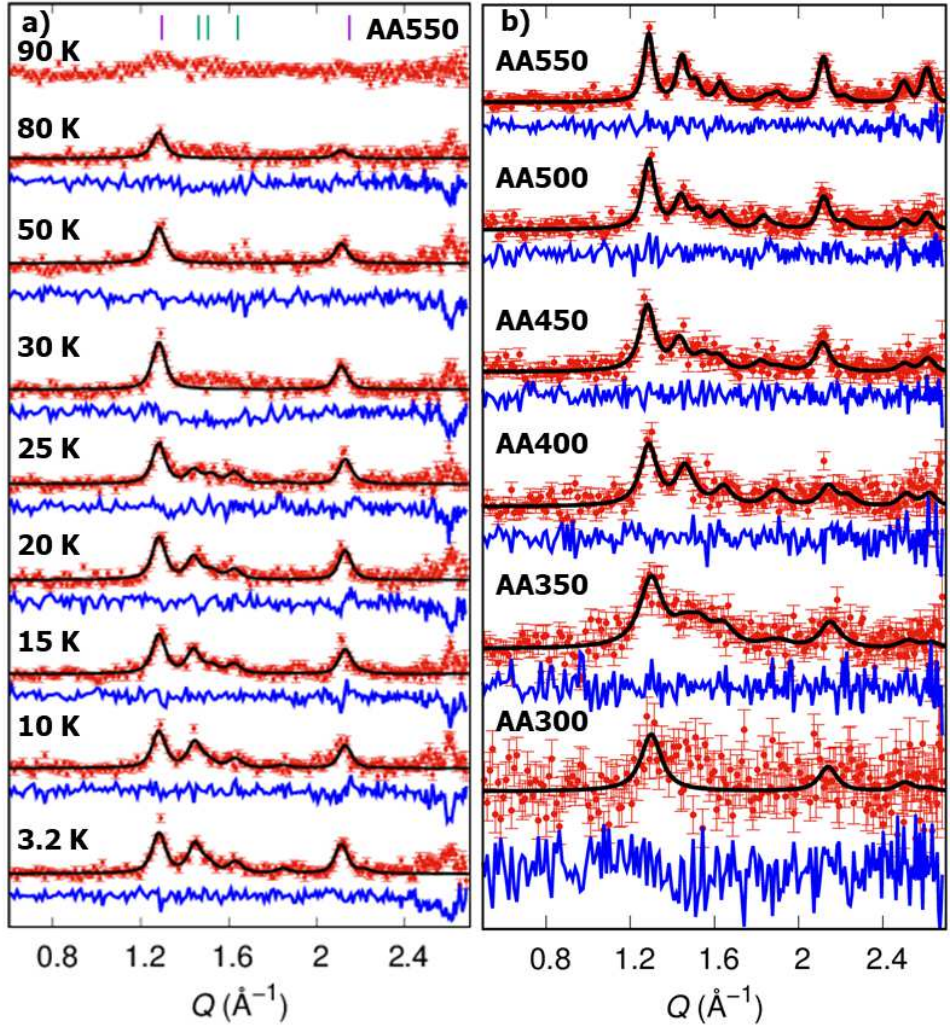}
	\caption{\textbf{a)} Temperature dependence of the magnetic scattering cross section (red dots) of cobalt chromite nanoparticle with coherent domain size of 14.0\,nm (sample AA550). The black line is Le-Bail fit and blue line is the residual between the fit and experimental data. Purple and green vertical lines represent the position of fundamental and satellite reflections, respectively. \textbf{b)} Magnetic scattering contribution of all studied samples recorded at 3.5\,K (red points: experimental data). The black line is Le-Bail fit and blue line is residual between the fit and experimental data.}
	\label{fig:PND}
\end{figure}

The magnetic scattering contribution (\textbf{Fig. \ref{fig:PND} a)}) directly relates to the magnetic phase structure. At the first magnetic phase transition, from paramagnetic to collinear ferrimagnetic magnetic ordering between 90 to 80\,K, fundamental magnetic reflections arise. The transition to noncollinear magnetic ordering around 25\,K is accompanied with additional magnetic satellite reflections. The magnetic reflections at 3.5\,K (\textbf{Fig. \ref{fig:PND} b)}) broaden significantly with decreasing coherent domain size (see \textbf{Fig S 6} for the complete temperature dependence of the magnetic scattering for all samples).

The temperature dependence of the integral reflection intensity is obtained from Le-Bail fits of the magnetic scattering cross sections, and is used as magnetic order parameter to determine the magnetic phase boundaries (\textbf{Fig. S 7} and \textbf{Fig. S 8})\cite{SupInf}. Particle size dependent magnetic phase transition temperatures are thus accessible with enhanced precision. The ferrimagnetic phase appears between 96(1) to 56(1)\,K, and the spin-spiral transition temperature changes from 28\,K (similar to bulk\cite{Yamasaki2006}) down to 3.5\,K with decreasing particle size (see \textbf{Table S III}\cite{SupInf}).

\begin{figure}[htpb]
	\centering
	\includegraphics[trim= 1cm 2cm 0 0,width=0.8\columnwidth]{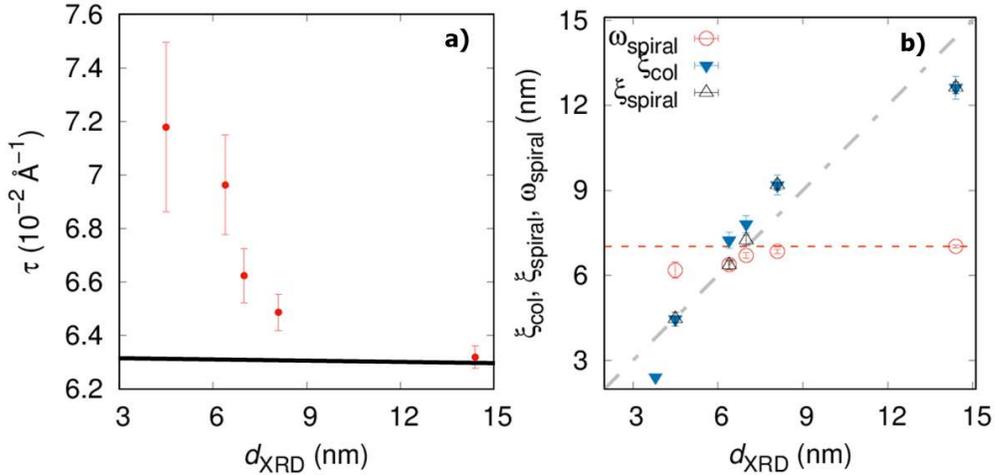}
	\caption{Size dependence of the \textbf{a)} propagation vector $\tau$ at the transition temperature (bulk value is presented by black line\cite{Zakutna2018}). \textbf{b)} Size dependence of the period of spin-spiral $\omega_\mathrm{spiral}$, and the coherence length of spin-spiral ordering $\xi_\mathrm{spiral}$ and collinear structure $\xi_\mathrm{col}$. The grey and red dashed lines represent $\xi$ = $d_\mathrm{XRD}$ and the bulk value of $\omega_\mathrm{spiral}$\cite{Chang2009}, respectively. }
	\label{fig:dependencies}
\end{figure}

Based on the satellite reflection in the magnetic neutron diffractograms, the spin-spiral propagation vector ($\tau\space\tau\space0$) was determined using a conical spin spiral model\cite{Hastings1962,Lyons1962}. In accordance with the method used to obtain the bulk reference\cite{Yamasaki2006}, the propagation vector was determined for each sample at the corresponding transition temperature $T_\mathrm{s}$. The propagation vector component $\tau$ slightly increases with decreasing particle size from $\tau$ = 6.32(1)$\cdot$10$^{-2}$ to 7.2(1)$\cdot$10$^{-2}$\,\AA$^{-1}$ (see \textbf{Fig. \ref{fig:dependencies} a)}). Correspondingly, the period of the spin spiral ($\omega_\mathrm{spiral} = 2\pi/\sqrt{2\tau^2}$) (\textbf{Fig. \ref{fig:dependencies} b)}) reduces together with the particle size. 
We observe that $\omega_\mathrm{spiral}$ approaches the particle size in the range 7 to 6.4 nm, indicating that exactly one period of the spin spiral is compressed slightly to fit into the NP. At lower particle size of 4.5\,nm, the spin spiral phase is still present, suggesting that less than one period may be accommodated within the NP. In case of the smallest particle size of 3.6 nm, the noncollinear phase is absent. The correlation lengths of the collinear ($\xi_\mathrm{col} = 2\pi/\mathrm{FWHM_\mathrm{col}}$, with FWHM the full-width at half maximum intensity) and non-collinear order ($\xi_\mathrm{spiral} = 2\pi/\mathrm{FWHM}_\mathrm{spiral}$) are in good agreement with the particle grain size(\textbf{Fig. \ref{fig:dependencies} b)}). For $d_\mathrm{XRD}$= 3.6\,nm (AA300) only a very broad fundamental magnetic reflection of a collinear magnetic state is observed at 3.5\,K. In conjunction with our prior study\cite{Zakutna2018}, NPs as small as 2.7\,nm reveal a magnetically frustrated, cluster glass behavior. We associate the diffuse magnetic scattering of the 3.6\,nm NPs with the crossover towards a short-range order.   

\begin{figure}[htpb]
	\centering
	\includegraphics[trim= 1cm 2cm 3cm 1cm,width=0.8\columnwidth]{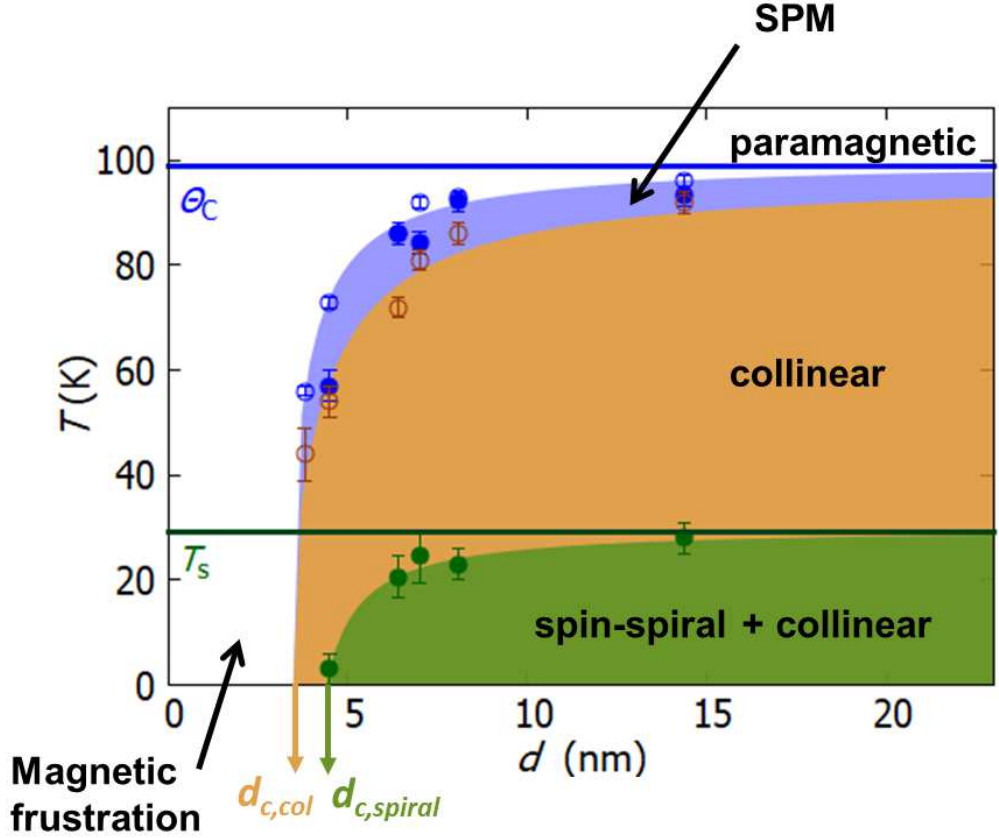}	\caption{Size-temperature dependence of the magnetic phase transitions. The blue and green horizontal lines in the magnetic phase diagram correspond to the bulk value of the Curie $\theta_\mathrm{c}$ and $T_\mathrm{s}$ spin-spiral transition temperature. The Curie $\theta_\mathrm{c}$ (blue), blocking $T_\mathrm{b}$ (brown) and spin-spiral transition $T_\mathrm{s}$ (green) temperatures are described by equation (\ref{eq1}). $d_\mathrm{c,spiral}$ and $d_\mathrm{c,col}$ denote the critical nanocrystalline size for the spin-spiral and ferrimagnetic ordering, respectively. Results obtained from magnetization and neutron diffraction measurements are described by open and full symbols, respectively.}
	\label{fig:phase}
\end{figure}

Combining polarized neutron diffraction with macroscopic magnetization results allows deriving the size induced magnetic phase evolution of nanocrystalline \ce{CoCr2O4} (\textbf{Fig. \ref{fig:phase}}). We describe the size dependence of the transition temperatures by the following equation\cite{Xie2005}:

\begin{equation}\label{eq1}
 	T(d) = T_\mathrm{bulk} (1 - \frac{C}{d}),
\end{equation}

where $T_\mathrm{bulk}$ is the respective transition temperature of the bulk material\cite{Yamasaki2006} and $C = 6\mu(6M)^{1/3}/(\rho\pi N_\mathrm{A})^{1/3}Z^{2/3}\pi k^{2}$ with $Z$ = 8 the formula unit, $k = \sqrt{2}/4$  the ratio between atomic radius and lattice parameter of fcc unit cell, $\mu$ shape factor $M$ the molecular weight, $N_\mathrm{A}$ Avogadro constant and $\rho$ mass density. All parameters in equation (1) were fixed, except the shape factor, which determines the curvature of the phase boundaries. For the bulk value of the blocking temperature, $T_\mathrm{b}$ = 94\,K was used, which corresponds to the onset of ferrimagnetic order in polycrystalline \ce{CoCr2O4}\cite{Lawes2006}. The 1/d dependence in equation (\ref{eq1}) indicates that the phase transition temperature is determined by the surface-to-volume fraction. We obtain critical particle sizes ($T(d_\mathrm{c})$ = 0) for the formation of the spin spiral magnetic structure $d_\mathrm{c,spiral}$ = 4.4(1)\,nm, for the collinear magnetic order $d_\mathrm{c,col}$ = 3.3(1)\,nm and for SPM behavior $d_\mathrm{c,spm}$ = 3.2(5)\,nm. Further reducing the particle size results in magnetic frustration and spin glass behavior as demonstrated for 2.7\,nm NPs in our previous work\cite{Zakutna2018}.

In addition to the magnetic properties, we find that the spin spiral magnetic order observed by neutron diffraction (see \textbf{Table S III}\cite{SupInf}) is accompanied with the onset of multiferroic behavior evidenced by macroscopic polarization measurements. The appearance of a ferroelectric transition at approximately 28\,K is clearly observed by means of higher harmonics measurements of the permittivity\cite{Niermann2014,Grams2019} for AA500 sample (see \textbf{Fig S 9}\cite{SupInf}). The transition temperature is in good agreement with $T_\mathrm{s} = 23(3)$\,K for this samples. As the NPs have a large surface-to-volume ratio, we expect contact effects to play an important role for direct measurements of the polarization with a high amount of organic surfactant hindering the measurement of a resolvable total macroscopic polarization signal. The incoherent scattering contribution in \textbf{Fig.S5}\cite{SupInf} is directly related to the amount of organic material, which varies strongly for each samples, with its lowest contribution in AA500. Considering that the onset of a ferroelectric signal was obtained only for the AA500 sample, we conclude the significant amount of organic material affects the polarization signal strongly for most samples. However, the clear signal obtained for AA500 establishes the multiferroic properties in \ce{CoCr2O4} NPs related to the spin spiral phase.\\

In summary, the particle size dependent magnetic phase diagram of nanocrystalline \ce{CoCr2O4} was elucidated by a combination of integral magnetization measurements and polarized neutron diffraction. The continuous decrease of the Curie, blocking and transition temperatures with nanocrystal size leads to a critical particle size of $d_\mathrm{c,spiral}$ = 4.4(1)\,nm, above which spin spiral magnetic order can exist. 
A minimum spin spiral period that can be squeezed entirely into the nanoparticle of 6.4(1)\,nm has been identified. Below a critical particle size of $d_\mathrm{c,col}$ = 3.3(1)\,nm, collinear magnetism is absent. Instead, the strong inter- and intraparticle interactions arising from structural and surface disorder in the \ce{CoCr2O4} NPs lead to a frustrated state known as collective super-spin glass behavior. 
Moreover, multiferroic behavior in nanocrystalline \ce{CoCr2O4} is confirmed for the first time. Low content of organic material is presumed crucial for successful dielectric spectroscopy. 
Our findings have direct consequences for potential applications of cobalt chromite NPs by defining the grain size limits for the presence of spin-spiral magnetic structure prerequisite for ferroelectricty in \ce{CoCr2O4}. Concerning the multiferroic properties, elimination of the organic residue in the NPs is important for enhancement of the ferroelectric signal and will pave the way towards the understanding of the coupling of multiferroic behavior with and spin spiral magnetic order in the \ce{CoCr2O4} NPs.

\begin{acknowledgements}
We thank Stefan Roitsch for TEM measurements. We thank Alena Klindziuk for her help with data evaluation in the frame of a GIANT internship at ILL in Grenoble, France. This work is based upon experiments performed at the DNS instrument operated by JCNS at the Heinz Maier-Leibnitz Zentrum (MLZ), Garching, Germany. Financial support from Ministry of Education, Youth and Sports of the Czech Republic in the frame of project Inter-COST LTC17058 and by COST Action CA15107 'MultiComp' and from the German Research Foundation (DFG: Emmy Noether Grant DI 1788/2-1 and project 277146847 – SFB 1238 (B02)) as well as the Institutional Strategy of the University of Cologne within the German Excellence Initiative (Max Delbr\"{u}ck-Prize for Young Researchers) are gratefully acknowledged. In Memoriam D. Ni\v{z}\v{n}ansk\'{y}.
\end{acknowledgements}

\bibliographystyle{apsrev}
\bibliography{references}

\begin{thebibliography}{27}
\expandafter\ifx\csname natexlab\endcsname\relax\def\natexlab#1{#1}\fi
\expandafter\ifx\csname bibnamefont\endcsname\relax
  \def\bibnamefont#1{#1}\fi
\expandafter\ifx\csname bibfnamefont\endcsname\relax
  \def\bibfnamefont#1{#1}\fi
\expandafter\ifx\csname citenamefont\endcsname\relax
  \def\citenamefont#1{#1}\fi
\expandafter\ifx\csname url\endcsname\relax
  \def\url#1{\texttt{#1}}\fi
\expandafter\ifx\csname urlprefix\endcsname\relax\def\urlprefix{URL }\fi
\providecommand{\bibinfo}[2]{#2}
\providecommand{\eprint}[2][]{\url{#2}}

\bibitem[{\citenamefont{Fiebig et~al.}(2016)\citenamefont{Fiebig, Lottermoser,
  Meier, and Trassin}}]{Fiebig2016}
\bibinfo{author}{\bibfnamefont{M.}~\bibnamefont{Fiebig}},
  \bibinfo{author}{\bibfnamefont{T.}~\bibnamefont{Lottermoser}},
  \bibinfo{author}{\bibfnamefont{D.}~\bibnamefont{Meier}}, \bibnamefont{and}
  \bibinfo{author}{\bibfnamefont{M.}~\bibnamefont{Trassin}},
  \bibinfo{journal}{Nature Reviews Materials} \textbf{\bibinfo{volume}{1}},
  \bibinfo{pages}{16046} (\bibinfo{year}{2016}).

\bibitem[{\citenamefont{Kimura et~al.}(2003)\citenamefont{Kimura, Ishihara,
  Shintani, Arima, Takahashi, Ishizaka, and Tokura}}]{Kimura2003}
\bibinfo{author}{\bibfnamefont{T.}~\bibnamefont{Kimura}},
  \bibinfo{author}{\bibfnamefont{S.}~\bibnamefont{Ishihara}},
  \bibinfo{author}{\bibfnamefont{H.}~\bibnamefont{Shintani}},
  \bibinfo{author}{\bibfnamefont{T.}~\bibnamefont{Arima}},
  \bibinfo{author}{\bibfnamefont{K.~T.} \bibnamefont{Takahashi}},
  \bibinfo{author}{\bibfnamefont{K.}~\bibnamefont{Ishizaka}}, \bibnamefont{and}
  \bibinfo{author}{\bibfnamefont{Y.}~\bibnamefont{Tokura}},
  \bibinfo{journal}{Phys. Rev. B} \textbf{\bibinfo{volume}{68}},
  \bibinfo{pages}{060403} (\bibinfo{year}{2003}).

\bibitem[{\citenamefont{Goto et~al.}(2004)\citenamefont{Goto, Kimura, Lawes,
  Ramirez, and Tokura}}]{Goto2004}
\bibinfo{author}{\bibfnamefont{T.}~\bibnamefont{Goto}},
  \bibinfo{author}{\bibfnamefont{T.}~\bibnamefont{Kimura}},
  \bibinfo{author}{\bibfnamefont{G.}~\bibnamefont{Lawes}},
  \bibinfo{author}{\bibfnamefont{A.~P.} \bibnamefont{Ramirez}},
  \bibnamefont{and} \bibinfo{author}{\bibfnamefont{Y.}~\bibnamefont{Tokura}},
  \bibinfo{journal}{Phys. Rev. Lett.} \textbf{\bibinfo{volume}{92}},
  \bibinfo{pages}{257201} (\bibinfo{year}{2004}).

\bibitem[{\citenamefont{Bousquet and Cano}(2016)}]{Bousquet2016}
\bibinfo{author}{\bibfnamefont{E.}~\bibnamefont{Bousquet}} \bibnamefont{and}
  \bibinfo{author}{\bibfnamefont{A.}~\bibnamefont{Cano}},
  \bibinfo{journal}{Journal of Physics: Condensed Matter}
  \textbf{\bibinfo{volume}{28}}, \bibinfo{pages}{123001}
  (\bibinfo{year}{2016}).

\bibitem[{\citenamefont{Higashiyama et~al.}(2004)\citenamefont{Higashiyama,
  Miyasaka, Kida, Arima, and Tokura}}]{Higashiyama2004}
\bibinfo{author}{\bibfnamefont{D.}~\bibnamefont{Higashiyama}},
  \bibinfo{author}{\bibfnamefont{S.}~\bibnamefont{Miyasaka}},
  \bibinfo{author}{\bibfnamefont{N.}~\bibnamefont{Kida}},
  \bibinfo{author}{\bibfnamefont{T.}~\bibnamefont{Arima}}, \bibnamefont{and}
  \bibinfo{author}{\bibfnamefont{Y.}~\bibnamefont{Tokura}},
  \bibinfo{journal}{Phys. Rev. B} \textbf{\bibinfo{volume}{70}},
  \bibinfo{pages}{174405} (\bibinfo{year}{2004}).

\bibitem[{\citenamefont{Mostovoy}(2006)}]{Mostovoy2006}
\bibinfo{author}{\bibfnamefont{M.}~\bibnamefont{Mostovoy}},
  \bibinfo{journal}{Phys. Rev. Lett.} \textbf{\bibinfo{volume}{96}},
  \bibinfo{pages}{067601} (\bibinfo{year}{2006}).

\bibitem[{\citenamefont{Cheong and Mostovoy}(2007)}]{Cheong2007}
\bibinfo{author}{\bibfnamefont{S.~W.} \bibnamefont{Cheong}} \bibnamefont{and}
  \bibinfo{author}{\bibfnamefont{M.}~\bibnamefont{Mostovoy}},
  \bibinfo{journal}{Nat Mater.} \textbf{\bibinfo{volume}{6}},
  \bibinfo{pages}{13} (\bibinfo{year}{2007}).

\bibitem[{\citenamefont{Yamasaki et~al.}(2006)\citenamefont{Yamasaki, Miyasaka,
  Kaneko, He, Arima, and Tokura}}]{Yamasaki2006}
\bibinfo{author}{\bibfnamefont{Y.}~\bibnamefont{Yamasaki}},
  \bibinfo{author}{\bibfnamefont{S.}~\bibnamefont{Miyasaka}},
  \bibinfo{author}{\bibfnamefont{Y.}~\bibnamefont{Kaneko}},
  \bibinfo{author}{\bibfnamefont{J.-P.} \bibnamefont{He}},
  \bibinfo{author}{\bibfnamefont{T.}~\bibnamefont{Arima}}, \bibnamefont{and}
  \bibinfo{author}{\bibfnamefont{Y.}~\bibnamefont{Tokura}},
  \bibinfo{journal}{Phys. Rev. Lett.} \textbf{\bibinfo{volume}{96}},
  \bibinfo{pages}{207204} (\bibinfo{year}{2006}).

\bibitem[{\citenamefont{Kim et~al.}(2009)\citenamefont{Kim, Seok~Oh, Liu, Chun,
  Lee, Ko, Park, Chung, and Kim}}]{Kim2009}
\bibinfo{author}{\bibfnamefont{I.}~\bibnamefont{Kim}},
  \bibinfo{author}{\bibfnamefont{Y.}~\bibnamefont{Seok~Oh}},
  \bibinfo{author}{\bibfnamefont{Y.}~\bibnamefont{Liu}},
  \bibinfo{author}{\bibfnamefont{S.}~\bibnamefont{Chun}},
  \bibinfo{author}{\bibfnamefont{J.-S.} \bibnamefont{Lee}},
  \bibinfo{author}{\bibfnamefont{K.-T.} \bibnamefont{Ko}},
  \bibinfo{author}{\bibfnamefont{J.-H.} \bibnamefont{Park}},
  \bibinfo{author}{\bibfnamefont{J.-H.} \bibnamefont{Chung}}, \bibnamefont{and}
  \bibinfo{author}{\bibfnamefont{K.~H.} \bibnamefont{Kim}},
  \bibinfo{journal}{Appl. Phys. Lett.} \textbf{\bibinfo{volume}{94}},
  \bibinfo{pages}{042505} (\bibinfo{year}{2009}).

\bibitem[{\citenamefont{Chang et~al.}(2009)\citenamefont{Chang, Huang, Li,
  Cheong, Ratcliff, and Lynn}}]{Chang2009}
\bibinfo{author}{\bibfnamefont{L.~J.} \bibnamefont{Chang}},
  \bibinfo{author}{\bibfnamefont{D.~J.} \bibnamefont{Huang}},
  \bibinfo{author}{\bibfnamefont{W.-H.} \bibnamefont{Li}},
  \bibinfo{author}{\bibfnamefont{S.-W.} \bibnamefont{Cheong}},
  \bibinfo{author}{\bibfnamefont{W.}~\bibnamefont{Ratcliff}}, \bibnamefont{and}
  \bibinfo{author}{\bibfnamefont{J.~W.} \bibnamefont{Lynn}},
  \bibinfo{journal}{Journal of Physics: Condensed Matter}
  \textbf{\bibinfo{volume}{21}}, \bibinfo{pages}{456008}
  (\bibinfo{year}{2009}).

\bibitem[{\citenamefont{Z{\'a}kutn{\'a}
  et~al.}(2014)\citenamefont{Z{\'a}kutn{\'a}, Repko, Matulkov{\'a}, Ni{\v z}{\v
  n}ansk{\'y}, Ardu, Cannas, Mantlikov{\'a}, and Vejpravov{\'a}}}]{Zakutna2014}
\bibinfo{author}{\bibfnamefont{D.}~\bibnamefont{Z{\'a}kutn{\'a}}},
  \bibinfo{author}{\bibfnamefont{A.}~\bibnamefont{Repko}},
  \bibinfo{author}{\bibfnamefont{I.}~\bibnamefont{Matulkov{\'a}}},
  \bibinfo{author}{\bibfnamefont{D.}~\bibnamefont{Ni{\v z}{\v n}ansk{\'y}}},
  \bibinfo{author}{\bibfnamefont{A.}~\bibnamefont{Ardu}},
  \bibinfo{author}{\bibfnamefont{C.}~\bibnamefont{Cannas}},
  \bibinfo{author}{\bibfnamefont{A.}~\bibnamefont{Mantlikov{\'a}}},
  \bibnamefont{and}
  \bibinfo{author}{\bibfnamefont{J.}~\bibnamefont{Vejpravov{\'a}}},
  \bibinfo{journal}{Journal of Nanoparticle Research}
  \textbf{\bibinfo{volume}{16}}, \bibinfo{pages}{2251} (\bibinfo{year}{2014}).

\bibitem[{\citenamefont{Z{\'a}kutn{\'a}
  et~al.}(2016)\citenamefont{Z{\'a}kutn{\'a}, Matulkov{\'a}, Kentzinger,
  Medlin, Su, Nemkovski, Disch, Vejpravov{\'a}, and Ni{\v z}{\v
  n}ansk{\'y}}}]{Zakutna2016}
\bibinfo{author}{\bibfnamefont{D.}~\bibnamefont{Z{\'a}kutn{\'a}}},
  \bibinfo{author}{\bibfnamefont{I.}~\bibnamefont{Matulkov{\'a}}},
  \bibinfo{author}{\bibfnamefont{E.}~\bibnamefont{Kentzinger}},
  \bibinfo{author}{\bibfnamefont{R.}~\bibnamefont{Medlin}},
  \bibinfo{author}{\bibfnamefont{Y.}~\bibnamefont{Su}},
  \bibinfo{author}{\bibfnamefont{K.}~\bibnamefont{Nemkovski}},
  \bibinfo{author}{\bibfnamefont{S.}~\bibnamefont{Disch}},
  \bibinfo{author}{\bibfnamefont{J.}~\bibnamefont{Vejpravov{\'a}}},
  \bibnamefont{and} \bibinfo{author}{\bibfnamefont{D.}~\bibnamefont{Ni{\v z}{\v
  n}ansk{\'y}}}, \bibinfo{journal}{RSC Adv.} \textbf{\bibinfo{volume}{6}},
  \bibinfo{pages}{107659} (\bibinfo{year}{2016}).

\bibitem[{\citenamefont{Z\'akutn\'a et~al.}(2018)\citenamefont{Z\'akutn\'a,
  Vl\ifmmode~\check{c}\else \v{c}\fi{}ek, Fitl, Nemkovski, Honecker, Ni\ifmmode
  \check{z}\else \v{z}\fi{}\ifmmode~\check{n}\else \v{n}\fi{}ansk\'y, and
  Disch}}]{Zakutna2018}
\bibinfo{author}{\bibfnamefont{D.}~\bibnamefont{Z\'akutn\'a}},
  \bibinfo{author}{\bibfnamefont{J.}~\bibnamefont{Vl\ifmmode~\check{c}\else
  \v{c}\fi{}ek}}, \bibinfo{author}{\bibfnamefont{P.}~\bibnamefont{Fitl}},
  \bibinfo{author}{\bibfnamefont{K.}~\bibnamefont{Nemkovski}},
  \bibinfo{author}{\bibfnamefont{D.}~\bibnamefont{Honecker}},
  \bibinfo{author}{\bibfnamefont{D.}~\bibnamefont{Ni\ifmmode \check{z}\else
  \v{z}\fi{}\ifmmode~\check{n}\else \v{n}\fi{}ansk\'y}}, \bibnamefont{and}
  \bibinfo{author}{\bibfnamefont{S.}~\bibnamefont{Disch}},
  \bibinfo{journal}{Phys. Rev. B} \textbf{\bibinfo{volume}{98}},
  \bibinfo{pages}{064407} (\bibinfo{year}{2018}).

\bibitem[{\citenamefont{Rath and Mohanty}(2011)}]{Rath2011}
\bibinfo{author}{\bibfnamefont{C.}~\bibnamefont{Rath}} \bibnamefont{and}
  \bibinfo{author}{\bibfnamefont{P.}~\bibnamefont{Mohanty}},
  \bibinfo{journal}{Journal of Superconductivity and Novel Magnetism}
  \textbf{\bibinfo{volume}{24}}, \bibinfo{pages}{629 } (\bibinfo{year}{2011}).

\bibitem[{\citenamefont{Dutta et~al.}(2009)\citenamefont{Dutta, Manjanna, and
  Tyagi}}]{Dutta2009}
\bibinfo{author}{\bibfnamefont{D.~P.} \bibnamefont{Dutta}},
  \bibinfo{author}{\bibfnamefont{J.}~\bibnamefont{Manjanna}}, \bibnamefont{and}
  \bibinfo{author}{\bibfnamefont{A.~K.} \bibnamefont{Tyagi}},
  \bibinfo{journal}{Journal of Applied Physics} \textbf{\bibinfo{volume}{106}},
  \bibinfo{pages}{043915} (\bibinfo{year}{2009}).

\bibitem[{\citenamefont{Kumar et~al.}(2009)\citenamefont{Kumar, Mohanty,
  Shripathi, and Rath}}]{Kumar2009}
\bibinfo{author}{\bibfnamefont{L.}~\bibnamefont{Kumar}},
  \bibinfo{author}{\bibfnamefont{P.}~\bibnamefont{Mohanty}},
  \bibinfo{author}{\bibfnamefont{T.}~\bibnamefont{Shripathi}},
  \bibnamefont{and} \bibinfo{author}{\bibfnamefont{C.}~\bibnamefont{Rath}},
  \bibinfo{journal}{Nanoscience and Nanotechnology Letters}
  \textbf{\bibinfo{volume}{1}}, \bibinfo{pages}{193 } (\bibinfo{year}{2009}).

\bibitem[{\citenamefont{Tian et~al.}(2015)\citenamefont{Tian, Zhu, Wang, Xia,
  Liu, and Yuan}}]{Tian2015}
\bibinfo{author}{\bibfnamefont{Z.}~\bibnamefont{Tian}},
  \bibinfo{author}{\bibfnamefont{C.}~\bibnamefont{Zhu}},
  \bibinfo{author}{\bibfnamefont{J.}~\bibnamefont{Wang}},
  \bibinfo{author}{\bibfnamefont{Z.}~\bibnamefont{Xia}},
  \bibinfo{author}{\bibfnamefont{Y.}~\bibnamefont{Liu}}, \bibnamefont{and}
  \bibinfo{author}{\bibfnamefont{S.}~\bibnamefont{Yuan}}, \bibinfo{journal}{J.
  Magn. Magn. Mater.} \textbf{\bibinfo{volume}{377}}, \bibinfo{pages}{176}
  (\bibinfo{year}{2015}).

\bibitem[{\citenamefont{Galivarapu
  et~al.}(2016{\natexlab{a}})\citenamefont{Galivarapu, Kumar, Banerjee, and
  Rath}}]{Galivarapu2016}
\bibinfo{author}{\bibfnamefont{J.~K.} \bibnamefont{Galivarapu}},
  \bibinfo{author}{\bibfnamefont{D.}~\bibnamefont{Kumar}},
  \bibinfo{author}{\bibfnamefont{A.}~\bibnamefont{Banerjee}}, \bibnamefont{and}
  \bibinfo{author}{\bibfnamefont{C.}~\bibnamefont{Rath}},
  \bibinfo{journal}{IEEE Transactions on Magnetics}
  \textbf{\bibinfo{volume}{52}}, \bibinfo{pages}{1}
  (\bibinfo{year}{2016}{\natexlab{a}}).

\bibitem[{\citenamefont{Galivarapu
  et~al.}(2016{\natexlab{b}})\citenamefont{Galivarapu, Kumar, Banerjee, Sathe,
  Aquilanti, and Rath}}]{GalivarapuRSC2016}
\bibinfo{author}{\bibfnamefont{J.~K.} \bibnamefont{Galivarapu}},
  \bibinfo{author}{\bibfnamefont{D.}~\bibnamefont{Kumar}},
  \bibinfo{author}{\bibfnamefont{A.}~\bibnamefont{Banerjee}},
  \bibinfo{author}{\bibfnamefont{V.}~\bibnamefont{Sathe}},
  \bibinfo{author}{\bibfnamefont{G.}~\bibnamefont{Aquilanti}},
  \bibnamefont{and} \bibinfo{author}{\bibfnamefont{C.}~\bibnamefont{Rath}},
  \bibinfo{journal}{RSC Adv.} \textbf{\bibinfo{volume}{6}},
  \bibinfo{pages}{63809} (\bibinfo{year}{2016}{\natexlab{b}}).

\bibitem[{\citenamefont{Kumar et~al.}(2016)\citenamefont{Kumar, Galivarapu,
  Banerjee, Nemkovski, Su, and Rath}}]{Kumar2016}
\bibinfo{author}{\bibfnamefont{D.}~\bibnamefont{Kumar}},
  \bibinfo{author}{\bibfnamefont{J.~K.} \bibnamefont{Galivarapu}},
  \bibinfo{author}{\bibfnamefont{A.}~\bibnamefont{Banerjee}},
  \bibinfo{author}{\bibfnamefont{K.~S.} \bibnamefont{Nemkovski}},
  \bibinfo{author}{\bibfnamefont{Y.}~\bibnamefont{Su}}, \bibnamefont{and}
  \bibinfo{author}{\bibfnamefont{C.}~\bibnamefont{Rath}},
  \bibinfo{journal}{Nanotechnology} \textbf{\bibinfo{volume}{27}},
  \bibinfo{pages}{175702} (\bibinfo{year}{2016}).

\bibitem[{Sup()}]{SupInf}
\bibinfo{note}{See Supplemental Material at [URL will be inserted by publisher]
  for experimental details, macroscopic magnetization measurements, results of
  the Le Bail analysis of PXRD data, a complete set of neutron diffractograms
  including temperature dependence of integral intensity of fundamental (111)
  and magnetic satellite (111(0)) reflections, and macroscopic polarization
  measurements.}

\bibitem[{\citenamefont{Hastings and Corliss}(1962)}]{Hastings1962}
\bibinfo{author}{\bibfnamefont{J.~M.} \bibnamefont{Hastings}} \bibnamefont{and}
  \bibinfo{author}{\bibfnamefont{L.~M.} \bibnamefont{Corliss}},
  \bibinfo{journal}{Physical Review} \textbf{\bibinfo{volume}{126}},
  \bibinfo{pages}{556 } (\bibinfo{year}{1962}).

\bibitem[{\citenamefont{Lyons et~al.}(1962)\citenamefont{Lyons, Kaplan, Dwight,
  and Menyuk}}]{Lyons1962}
\bibinfo{author}{\bibfnamefont{D.~H.} \bibnamefont{Lyons}},
  \bibinfo{author}{\bibfnamefont{T.~A.} \bibnamefont{Kaplan}},
  \bibinfo{author}{\bibfnamefont{K.}~\bibnamefont{Dwight}}, \bibnamefont{and}
  \bibinfo{author}{\bibfnamefont{N.}~\bibnamefont{Menyuk}},
  \bibinfo{journal}{Physical Review} \textbf{\bibinfo{volume}{126}},
  \bibinfo{pages}{540 } (\bibinfo{year}{1962}).

\bibitem[{\citenamefont{Xie et~al.}(2005)\citenamefont{Xie, Wang, and
  Cao}}]{Xie2005}
\bibinfo{author}{\bibfnamefont{D.}~\bibnamefont{Xie}},
  \bibinfo{author}{\bibfnamefont{M.}~\bibnamefont{Wang}}, \bibnamefont{and}
  \bibinfo{author}{\bibfnamefont{L.}~\bibnamefont{Cao}},
  \bibinfo{journal}{physica status solidi (b)} \textbf{\bibinfo{volume}{242}},
  \bibinfo{pages}{R76} (\bibinfo{year}{2005}).

\bibitem[{\citenamefont{Lawes et~al.}(2006)\citenamefont{Lawes, Melot, Page,
  Ederer, Hayward, Proffen, and Seshadri}}]{Lawes2006}
\bibinfo{author}{\bibfnamefont{G.}~\bibnamefont{Lawes}},
  \bibinfo{author}{\bibfnamefont{B.}~\bibnamefont{Melot}},
  \bibinfo{author}{\bibfnamefont{K.}~\bibnamefont{Page}},
  \bibinfo{author}{\bibfnamefont{C.}~\bibnamefont{Ederer}},
  \bibinfo{author}{\bibfnamefont{M.~A.} \bibnamefont{Hayward}},
  \bibinfo{author}{\bibfnamefont{T.}~\bibnamefont{Proffen}}, \bibnamefont{and}
  \bibinfo{author}{\bibfnamefont{R.}~\bibnamefont{Seshadri}},
  \bibinfo{journal}{Phys. Rev. B} \textbf{\bibinfo{volume}{74}},
  \bibinfo{pages}{024413} (\bibinfo{year}{2006}).

\bibitem[{\citenamefont{Niermann et~al.}(2014)\citenamefont{Niermann, Grams,
  Schalenbach, Becker, Bohat\'y, Stein, Braden, and Hemberger}}]{Niermann2014}
\bibinfo{author}{\bibfnamefont{D.}~\bibnamefont{Niermann}},
  \bibinfo{author}{\bibfnamefont{C.}~\bibnamefont{Grams}},
  \bibinfo{author}{\bibfnamefont{M.}~\bibnamefont{Schalenbach}},
  \bibinfo{author}{\bibfnamefont{P.}~\bibnamefont{Becker}},
  \bibinfo{author}{\bibfnamefont{L.}~\bibnamefont{Bohat\'y}},
  \bibinfo{author}{\bibfnamefont{J.}~\bibnamefont{Stein}},
  \bibinfo{author}{\bibfnamefont{M.}~\bibnamefont{Braden}}, \bibnamefont{and}
  \bibinfo{author}{\bibfnamefont{J.}~\bibnamefont{Hemberger}},
  \bibinfo{journal}{Phys. Rev. B} \textbf{\bibinfo{volume}{87}},
  \bibinfo{pages}{134412} (\bibinfo{year}{2014}).

\bibitem[{\citenamefont{Grams et~al.}(2019)\citenamefont{Grams, Kopatz,
  Br\"{u}ning, Biesenkamp, Becker, Bohat\'y, Lorenz, and
  Hemberger}}]{Grams2019}
\bibinfo{author}{\bibfnamefont{C.}~\bibnamefont{Grams}},
  \bibinfo{author}{\bibfnamefont{S.}~\bibnamefont{Kopatz}},
  \bibinfo{author}{\bibfnamefont{D.}~\bibnamefont{Br\"{u}ning}},
  \bibinfo{author}{\bibfnamefont{S.}~\bibnamefont{Biesenkamp}},
  \bibinfo{author}{\bibfnamefont{P.}~\bibnamefont{Becker}},
  \bibinfo{author}{\bibfnamefont{L.}~\bibnamefont{Bohat\'y}},
  \bibinfo{author}{\bibfnamefont{T.}~\bibnamefont{Lorenz}}, \bibnamefont{and}
  \bibinfo{author}{\bibfnamefont{J.}~\bibnamefont{Hemberger}},
  \bibinfo{journal}{Sci. Rep.} \textbf{\bibinfo{volume}{9}},
  \bibinfo{pages}{4391} (\bibinfo{year}{2019}).

\end{thebibliography}

\end{document}